\documentclass[smallextended]{svjour3}  %

\usepackage{graphicx}  
\usepackage{dcolumn}   
\usepackage{bm}        
\usepackage{amssymb}   
\usepackage{epstopdf}
\journalname{General Relativity and Gravitation}

\begin{document}
\renewcommand{\vec}[1]{\mathbf{#1}}


\title{Spinning solutions in general relativity with infinite central density}
%
%
\author{P.D.~Flammer} 
\institute{P.D.~Flammer \at
  Colorado School of Mines, Golden, Colorado, USA
  \email{pflammer@mines.edu}
}

\date{Received: Oct. 28, 2017 / Accepted: Mar. 31, 2018}

\maketitle

\begin{abstract}
This paper presents general relativistic numerical simulations of uniformly rotating polytropes. Equations are developed using MSQI coordinates, but taking a logarithm of the radial coordinate. The result is relatively simple elliptical differential equations. Due to the logarithmic scale, we can resolve solutions with near-singular mass distributions near their center, while the solution domain extends many orders of magnitude larger than the radius of the distribution (to connect with flat space-time). Rotating solutions are found with very high central energy densities for a range of adiabatic exponents. Analytically, assuming the pressure is proportional to the energy density (which is true for polytropes in the limit of large energy density), we determine the small radius behavior of the metric potentials and energy density. This small radius behavior agrees well with the small radius behavior of large central density numerical results, lending confidence to our numerical approach. We compare results with rotating solutions available in the literature, which show good agreement. We study the stability of spherical solutions: instability sets in at the first maximum in mass versus central energy density; this is also consistent with results in the literature, and further lends confidence to the numerical approach.
\end{abstract}

\section{Introduction}
Neutron stars consist of the densest naturally occurring material known in the universe, requiring general relativity to describe their dynamics. Numerical solutions for rotating neutron stars have been studied extensively in the literature\cite{friedman1986,stergioulas1995}. Komatsu et al. studied polytropes both uniformly and differentially rotating\cite{komatsu1989a,komatsu1989b}.  Cook et al. simulated various equations of state, including polytropes, for a range of central energy densities, and discovered the phenomenon of star ``spin-up'': an increase in rotation frequency as the star loses angular momentum and approaches instability\cite{cook1992,cook1994,cook1994b}. Bonazzola, Gourgoulhon, and others developed high precision models of rotating stars using maximally-sliced quasi-isotropic (MSQI) coordinates\cite{bonazzola1993,bonazzola1994,salgado1994}. Various other authors have extended upon these methods in the literature. See Refs.~\cite{stergioulas2003,gourgoulhon2010,shibata} and the references therein for an extensive review of the subject.

Many static solutions in general relativity are unstable to perturbations\cite{friedman2011}. In the static solution space, while following a contour of constant angular momentum, if the change in mass with respect to increasing central energy density ($dM/d\epsilon_0$) is less than zero, those solutions are secularly unstable against axisymmetric perturbations\cite{friedman1988,friedman1996,baumgarte2000}. A maximum of $M(\epsilon_0)$ on a contour is often used to determine the onset of instability; however, it has been shown that while this is true for non-rotating solutions, for rotating solutions, instability sets in at slightly smaller $\epsilon_0$\cite{takami2011}. For spherical distributions, a sufficient condition for instability due to radial perturbations was also reported by Chandrasekhar\cite{chandrasekhar1964prl,chandrasekhar1964}.

For rapidly rotating situations, non-axisymmetric instabilities (bar-mode formation) can occur. These occur on a secular time scale if the kinetic energy, $T$, is a significant fraction of the gravitational binding energy $W$, with the approximate condition $T/|W|>0.14$. Bar-mode instabilities set in on a dynamical time scale given the approximate condition $T/|W|>0.25$\cite{baumgarte2000,friedman1996,saijo2001,baiotti2007,manca2007}.

Spherical (non-rotating) solutions with extremely large central energy density, apparently limiting to infinite central energy density have been reported\cite{misner1964,misner1964Errata}. Although the sign of $dM/d\epsilon_0$ oscillates between positive and negative for higher energy densities, the distributions were found to be unstable using Chandrasekhar's instability condition\cite{chandrasekhar1964prl,misner1964}: spherical distributions are only stable out to the first maximum in $M(\epsilon_0)$; higher central energy densities are unstable. Such high central energy density solutions, but which are rotating, are the subject of this study.

We shall develop general relativistic equations for axisymmetric, time-independent situations, which are well suited to mass distributions with extremely high central energy densities. MSQI coordinates are used, but before solving, we take the logarithm of the radial coordinate. The resulting equations are then solved numerically using a finite element discretization.

As an internal check that the solver is working correclty, the 2D Virial theorem (GRV2)\cite{bonazzola1994} is used. As an external check, various solutions from our solver were compared to solutions using the freely available package LORENE. For the reader, we present a comparison of our results to results already in the literature (in Fig.~\ref{fig:mPlots}). We find our results in good agreement with those from LORENE and in the literature.

Finally, using Einstein's equations in spherically symmetric coordinates (in the radial gauge), we analytically determine the form of the energy density and metric components in the limit that the pressure is proportional to the energy density (which is true for polytropes of very high energy density). This agrees well with the small radius behavior of the highest central energy density numerical solutions; this also supports our numerical approach.

\section{Equations and Discretization}
We restrict ourselves to stationary states, which have cylindrical symmetry (axisymmetric). Additionally, we assume that all currents are circular (no meridional currents).

Using quasi-isotropic spherical polar coordinates $(t,r,\theta,\phi)$, with $g_{r\theta}=0$ and $g_{\theta\theta}=r^2g_{r r}$, the metric is parametrized as
\begin{equation}
\begin{array}{lll}
g_{\alpha\beta}d x^\alpha d x^\beta&=&-e^{2\nu} (c d t)^2+e^{2\zeta-2\nu}(d r^2+r^2d\theta^2)\\
&+&e^{2\gamma-2\nu}r^2\sin^2\theta\left[d\phi-\frac{1}{r_0}\omega (c d t)\right]^2,
\end{array}
\label{bonG}
\end{equation}
where $r_0$ is an arbitrary constant with units of length, which will provide a length scale to the problem, and $c$ is the speed of light. Our (unitless) metric potentials, $\nu$, $\gamma$, $\zeta$, and $\omega$, are functions of only $r$ and $\theta$. To develop equations of motion, we use the 3+1 formalism foliated using maximal slicing (a trace free extrinsic curvature tensor, $K=0$). With the definitions, $\gamma\equiv\ln(G)$, $\omega\equiv r_0 N^\phi$, this is the same formulation used by Bonazzola et al.\cite{bonazzola1993} (see Ref. \cite{bonazzola1993} for descriptions of $G$ and $N^\phi$).

In order to deal with near-singular mass distributions at the origin (or distributions that vary rapidly near the origin), we transform the radial coordinate using a logarithm, 
\begin{equation}
s\equiv\ln(r/r_0).
\label{logR}
\end{equation}
With this, Einstein's equations take on a relatively simple quasi-elliptical form, in terms of flat-space derivatives:
\begin{equation}
\begin{array}{lcl}
\nabla\cdot\left(
\begin{array}{c}
\partial_s \zeta\\
\partial_\theta \zeta
\end{array}
\right)&=&F_\zeta\\
\nabla\cdot\left(
\begin{array}{c}
\nu+\partial_s\nu\\
\partial_\theta \nu
\end{array}
\right)&=&F_\nu\\
\nabla\cdot\left(
\begin{array}{c}
2\gamma+\partial_s\gamma\\
\partial_\theta \gamma
\end{array}
\right)&=&F_\gamma\\
\nabla\cdot\left(
\begin{array}{c}
3\omega+\partial_s \omega\\
\partial_\theta \omega
\end{array}
\right)&=&F_\omega\\
\end{array}
\label{metricEquations}
\end{equation}
\begin{equation}
\begin{array}{rcl}
F_\zeta&\equiv&\sigma_{\rm fields}+\lambda \sigma_{\rm quad}\\
\sigma_{\rm fields}&\equiv&\frac{8 \pi}{\sin^2\theta} e^{2 \zeta-2 \gamma} T_{\phi\phi}\\
\sigma_{\rm quad}&\equiv&-\partial(\nu)\partial(\nu) +\frac{3}{4}\sin^2\theta e^{2 (\gamma - 2 \nu + s)} \partial(\omega)\partial(\omega)
\end{array}
\label{fZeta}
\end{equation}
\begin{equation}
\begin{array}{rcl}
F_\nu&\equiv&-\partial(\gamma)\partial(\nu)-\cot\theta \partial_\theta\nu\\
&+&\frac{1}{2}\sin ^2\theta e^{2 \gamma-4 \nu+2 s} \partial(\omega)\partial(\omega)\\
&+&4 \pi(r_0^2 e^{2 s} T_{r r} + T_{\theta\theta} + \frac{e^{2 \zeta-2 \gamma}}{\sin^2\theta}T_{\phi\phi})\\
&+&4 \pi e^{-4 \nu+2 s + 2 \zeta}(r_0^2 T_{tt} + 2 \omega r_0 T_{t\phi}+ \omega^2 T_{\phi\phi})
\end{array}
\end{equation}
\begin{equation}
\begin{array}{rcl}
F_\gamma&\equiv&-\partial(\gamma)\partial(\gamma) - 2 \cot\theta \partial_\theta\gamma + 8 \pi (T_{\theta\theta} + r_0^2 e^{2 s} T_{r r})
\end{array}
\end{equation}
\begin{equation}
\begin{array}{rcl}
F_\omega&\equiv&-3\partial(\omega)\partial(\gamma) + 4 \partial(\omega)\partial(\nu) - 3 \cot\theta \partial_\theta\omega\\
&+&\frac{16 \pi}{\sin^2\theta}e^{-2\gamma + 2\zeta}(\omega T_{\phi\phi} + r_0 T_{t\phi}).
\end{array}
\label{metricF}
\end{equation}
$T_{\alpha\beta}$ is the stress-energy tensor (note that appropriate factors of $G$, the gravitational constant, and $c$ are folded into $T_{\alpha\beta}$, or equivalently, one may say $G=c=1$ as is common); $\partial_s$ and $\partial_\theta$ are the partial derivatives with respect to $s$ and $\theta$; $\nabla$ is the flat-space gradient operator $\nabla=(\partial_s,\partial_\theta)$; and $\partial(f)\partial(g)\equiv\partial_s f\partial_s g + \partial_\theta f \partial_\theta g$, where $f$ and $g$ are any two functions.

$\lambda$ is an artificial addition to Eq.~\ref{fZeta}: for any true solution, $\lambda=1$. It is introduced into the equations in order to guarantee, as the solver converges to a solution, that $\int F_\zeta d s d\theta=0$. If this integral were non-zero, $\zeta$ would diverge linearly as a function of $s$ as $s\rightarrow\infty$ (or logarithmically as a function of $r$). Therefore, $\lambda$ is considered an unknown, and is solved for, by constraining $\int F_\zeta ds d\theta = 0$ (this is the 2D Virial Theorem). Checking how much $\lambda$ differs from 1 can be useful for identifying how accurate a solution is\cite{bonazzola1993}.

For this paper, we restrict ourselves to simulating a perfect polytropic fluid\cite{komatsu1989a},
\begin{equation}
\begin{array}{rcl}
T^{\alpha\beta}&=&(\epsilon+p)u^\alpha u^\beta+p g^{\alpha\beta}\\
p&=&K n^\Gamma,\\
\epsilon=\rho+\frac{1}{\Gamma-1}p&=&K\left(K_2 n+\frac{1}{\Gamma-1}n^\Gamma\right),
\end{array}
\label{polytrope}
\end{equation}
where $u^\alpha=(u^t,0,0,u^\phi)$ is the 4-velocity of the fluid, $p$ is the pressure, $\Gamma$ is the adiabatic exponent (and is constant), $\epsilon$ is the energy density, and $\rho=K K_2 n$ is the rest-mass density\cite{stergioulas2003}. The positions of the constants in Eq.~\ref{polytrope} differ slightly from what is commonly found in the literature (but is equivalent): we have rearranged constants so $K_2$ and $n$ are both unitless, while $K$ has units of 1/distance$^2$; this consolidates all units into $K$.

All physical variables presented here, such as the mass or radius of a distribution, are made unitless by multiplying by appropriate powers of $K$\cite{cook1994}. 

Imposing conservation of the stress-energy tensor, a first integral of motion is obtained in terms of the log enthalpy, $H$\cite{bonazzola1993}:
\begin{equation}
\begin{array}{rcl}
H&=&H_0-(\nu-\nu_0)+\ln(u^\alpha \nabla_\alpha t)-\int_{\Omega_0}^\Omega u_\phi u^t d\Omega
\end{array}
\label{firstIntegral}
\end{equation}
where $\nabla_\alpha$ is the covariant derivative, $\Omega$ is the angular velocity, and the subscript 0 signifies evaluation at $r=r_{\rm min}$, $\theta=\pi/2$.

$H$ is defined by its relationship to $n$ as\cite{bonazzola1993}
\begin{equation}
\begin{array}{rcl}
n&=&\left(\frac{\Gamma-1}{\Gamma}K_2(e^H-1)\right)^{\frac{1}{\Gamma-1}}.
\end{array}
\label{nOfH}
\end{equation}
Restricting ourselves to uniform rotation, the angular frequency, $\Omega$, is equal to the central angular frequency, $\Omega_0$, everywhere, and the last term in Eq.~\ref{firstIntegral} vanishes. 

Setting the arbitrary constant $r_0$ from Eq.~\ref{bonG} to $1/\sqrt{K}$ cancels all occurrences of both $K$ and $r_0$ in the differential equations. All the solutions presented here are therefore applicable for any $K$. To match other results in the literature\cite{cook1992,cook1994,cook1994b}, $K_2=1$ for all results presented here.

These equations are typically only solvable with numerical techniques. We choose finite elements to discretize the system of equations, using the commercially available package, Comsol Multiphysics. Since the domain of $s$ extends from $-\infty$ to $\infty$, we must artificially truncate it. The minimum radius, $r_{\rm min}$, is chosen such that the smallness of the volume element for $r<r_{\rm min}$ makes that region negligible (even for energy densities approaching infinity). The maximum distance simulated, $r_{\rm max}$, is chosen so all of the potentials are sufficiently close to 0 to approximate asymptotically flat space. For all simulations presented here, $r_{\rm min}=r_0\times10^{-8}$ ($s_{\rm min}\approx-18.4$) and $r_{\rm max}=r_0\times10^8$ ($s_{\rm max}\approx18.4$).

The boundary conditions used are: at $\theta=0,\pi$, the $\theta$ derivative of all metric potentials are zero; at $r_{\rm max}$, all metric potentials are zero (flat space-time); at $r_{\rm min}$, the normal derivative of all metric potentials are zero (regularity condition)\footnote{Even without regularity, for any function that has a finite derivative with respect to $r$ at $r=0$, the derivative with respect to $s$ will limit to $0$ as $r\rightarrow 0$, $s\rightarrow -\infty$. This is clear from the relation $\frac{\partial f}{\partial s} = \frac{\partial f}{\partial r}\frac{dr}{ds} = r_0 e^s \frac{\partial f}{\partial r} = r \frac{\partial f}{\partial r}$}.

The terms in the differential equations containing $\cot\theta \partial_\theta f $, where $f$ is a metric potential, are numerically troublesome at the poles; however, they are not truly singular as the $\theta$ derivative of each potential is zero at the poles. Those terms are approximated near the poles by tailor expanding $f$ and $\sin\theta$ about the pole; for instance near $\theta=0$, $\cot\theta \partial_\theta f \approx \cos\theta\partial_{\theta\theta} f/(1-\theta^2/6)$. The $1/\sin^2\theta$ in certain terms in the differential equations are canceled by factors of $\sin^2\theta$ in $T_{\alpha\beta}$.

To verify these equations, they were algebraically compared, using a commercial algebraic manipulation tool, Mathematica, to those of Ref.~\cite{bonazzola1993} (transforming $s\rightarrow r$ and transforming certain potentials), and were found to be identical. Additionally, various spherically symmetric and uniformly rotating models were compared to simulations using LORENE with the same parameters, with good agreement (relative error of less than $10^{-6}$ in the mass and equatorial radius for all models tested).

For an internal check as the numerical solver converged to a solution, we monitored the 2D Virial identity, GRV2\cite{bonazzola1994}. In terms of our equations, GRV2 is: $\int F_\zeta ds d\theta = 0$ with $\lambda=1$. While solving, we forced the Virial identity to be true and solved for $\lambda$. Monitoring how close $\lambda$ comes to converging to 1 has been shown to be a good indicator of the accuracy of solutions\cite{bonazzola1994}. In all cases, as our solver converged to a solution, $\lambda$ approached 1, and in cases where the solver failed to find a solution,  $\lambda$ failed to converge to values near 1. Increasing the mesh density also improved the convergence to 1. For simulations presented here, $\log(\lambda-1)$ was always less than -4, and typically less than -6. These results support the numerical solver is working correctly, and errors are primarily due to discretization.

We directly compared results of our simulation to those available in the literature. Fig.~\ref{fig:mPlots}(b) shows a direct comparison of the results from our simulation to those of Ref.~\cite{cook1994} Fig.~10. The two results are indistinguishable for the entire solution domain of Ref.~\cite{cook1994}.

In the Results section, we also compare the small-radius behavior of high central energy density solutions to analytic solutions of Einstein's equations. These are also in good agreement.

All of these results give confidence the numerical solver is working correctly, and is producing accurate solutions. 

\section{Results}
\begin{figure}
{
\begin{center}
\includegraphics[scale=0.6]{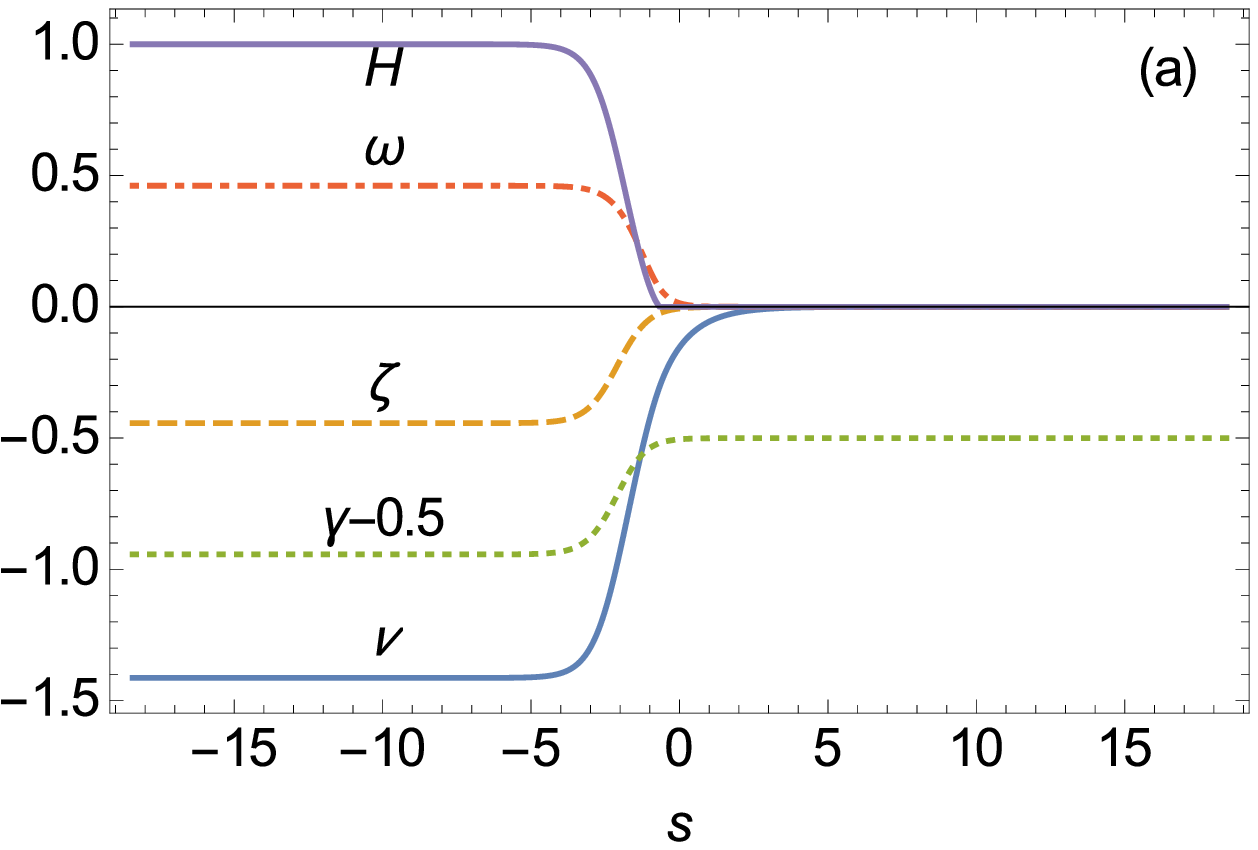}
\includegraphics[scale=0.6]{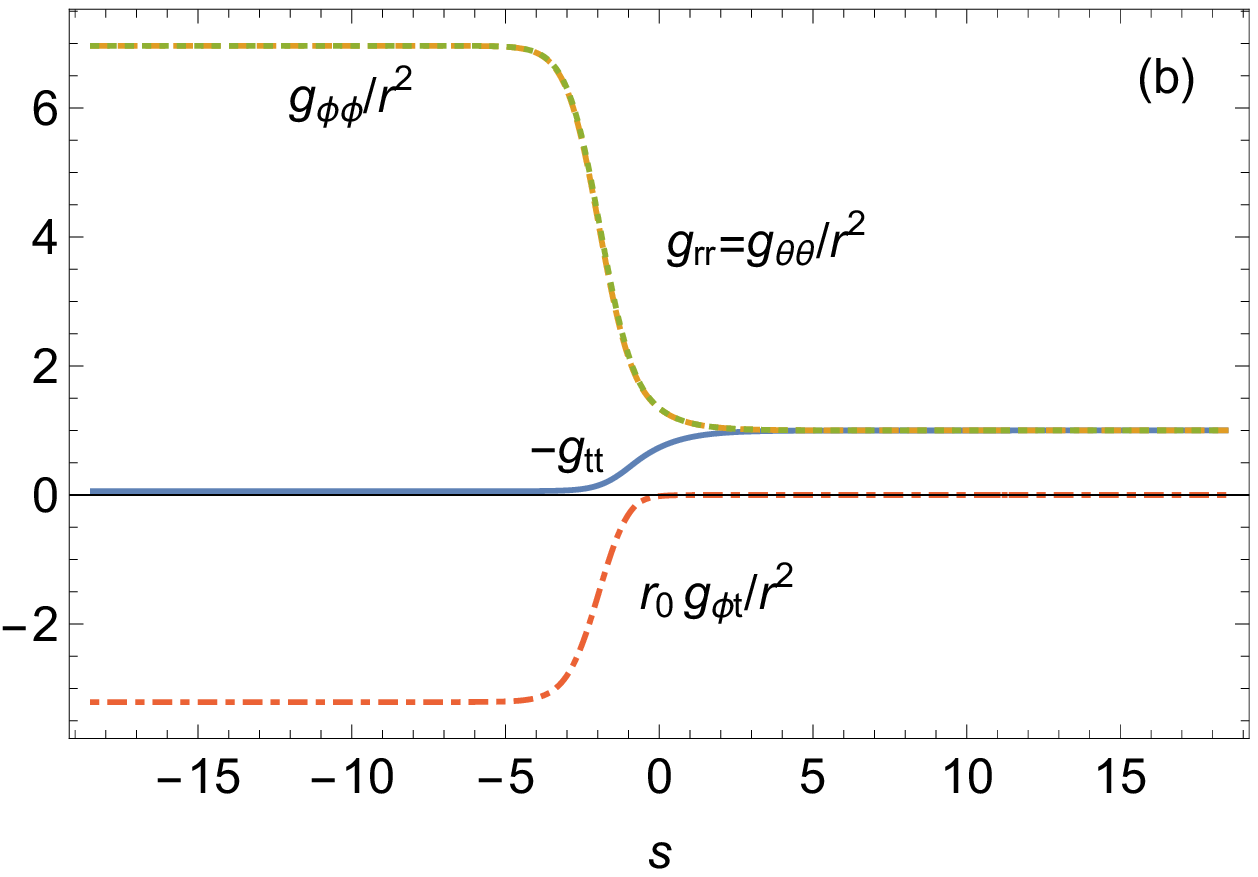}
\end{center}
\caption{\label{fig:typicalSoln}(a) The metric potentials and $H$, and (b) the metric coefficients as a function of $s$ on the equatorial plane. $\gamma$ is shifted by $0.5$ to differentiate it from $\zeta$ (they only differ slightly). $g_{\phi\phi}$ and $g_{\theta\theta}$ are slightly different, but are indistinguishable in this plot. For this solution, $\Gamma=2$, $H_0=1.0$, and $\Omega_0=0.55$.}
}
\end{figure}

The metric potentials, $H$, and the metric coefficients as a function of $s$ on the equatorial plane are shown for a typical solution in Fig.~\ref{fig:typicalSoln}. This solution uses an adiabatic exponent, $\Gamma=2$, has a central enthalpy, $H_0=1.0$, and angular frequency, $\Omega_0=0.55$; this is close to the maximum allowed (mass-shedding) angular frequency of 0.58. 

\begin{figure}
{
\begin{center}
\includegraphics[scale=0.6]{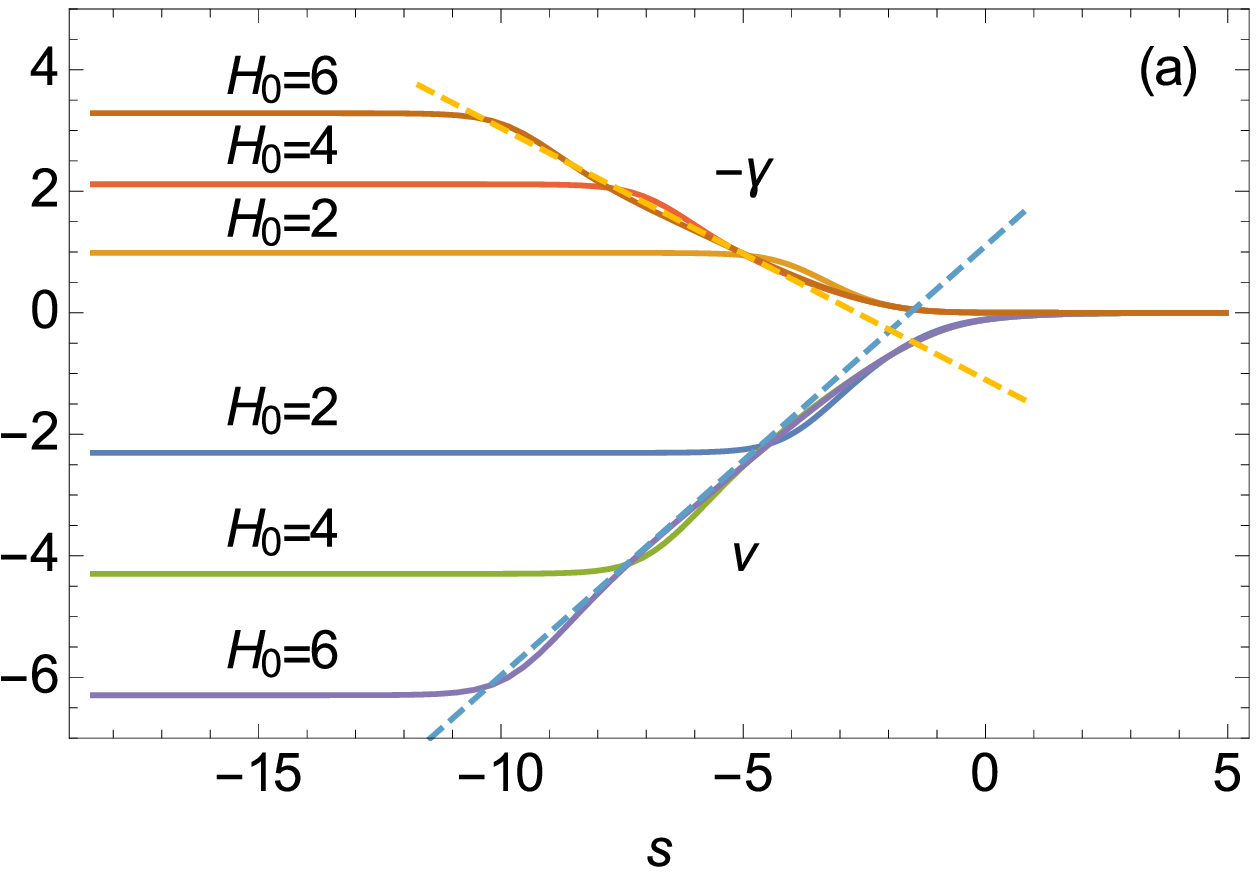}
\includegraphics[scale=0.6]{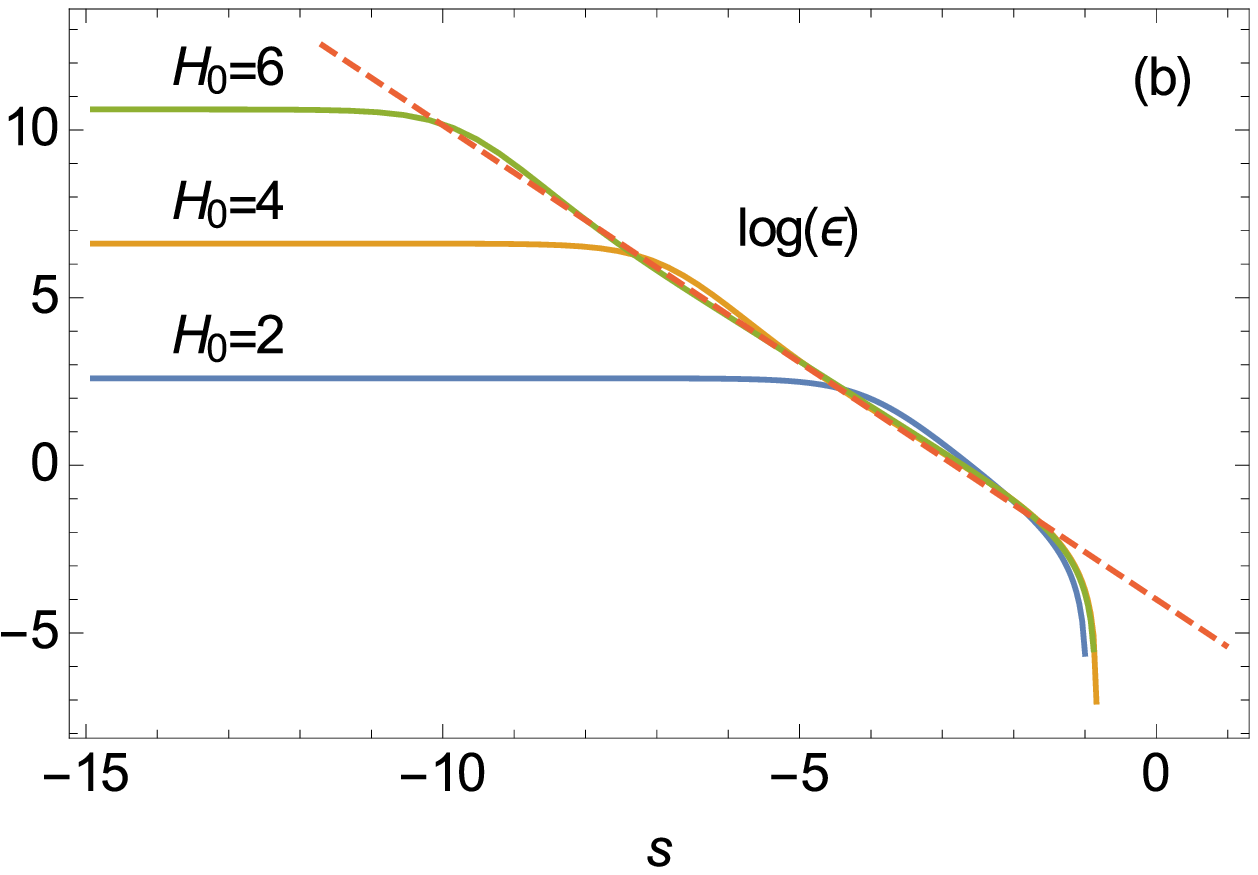}
\end{center}
\caption{\label{fig:trendNearOrigin}For $\Gamma=2$ and $\Omega_0=0.5$, how (a) $\gamma$ and $\nu$, and (b) $\log(\epsilon)$, trend as the central enthalpy $H_0$ is increased. For reference, the maximum $\Omega_0$ (mass-shedding limit) is: 0.596 for $H_0=2$, 0.511 for $H_0=4$, and 0.518 for $H_0=6$; so these correspond to very rapid rotations. The trend for large $H_0$ is not changed significantly by changing $\Omega_0$. The dashed curves are the analytic predictions of Eqs.~\ref{nuLargeE}-\ref{eLargeE}, showing good agreement with the numerical simulation.}
}
\end{figure}

Solutions are uniquely defined by $\Omega_0$, $H_0$. The allowed $\Omega_0$-$H_0$ space of static solutions does not have any upper bound on $H_0$. Fig.~\ref{fig:trendNearOrigin} shows, with $\Gamma=2$ and $\Omega_0=0.5$, how $\log(\epsilon)$, $\nu$ and $\gamma$ trend as $H_0$ increases. For very large central energy density, the uniform rotation has little effect on the small-$s$ (small-$r$) behavior of the energy density and metric potentials (outside of $\omega$), even near the mass-shedding limit.

\begin{figure}
{
\begin{center}
\includegraphics[scale=0.6]{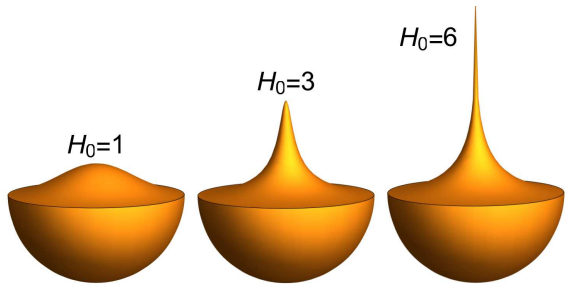}
\end{center}
\caption{\label{fig:3dPlots}(a) $H$ as a function of $r$ as $H_0$ is increased for $\Gamma=2$ and $\Omega_0=0$. The hemispheres represent the boundary of the mass distribution, which is changed little by the internal differences in solutions.}
}
\end{figure}

To give some perspective of how these solutions trend in normal (non-logarithmic) radial coordinates, Fig.~\ref{fig:3dPlots} shows $H$ as a function of $r$ as $H_0$ is increased. This demonstraties the singular nature of these solutions; note the actual energy is significantly more singular than shown here, as $\epsilon$ is proportional to $\exp(H)$. 

\begin{figure}
{
\begin{center}
\includegraphics[scale=0.45]{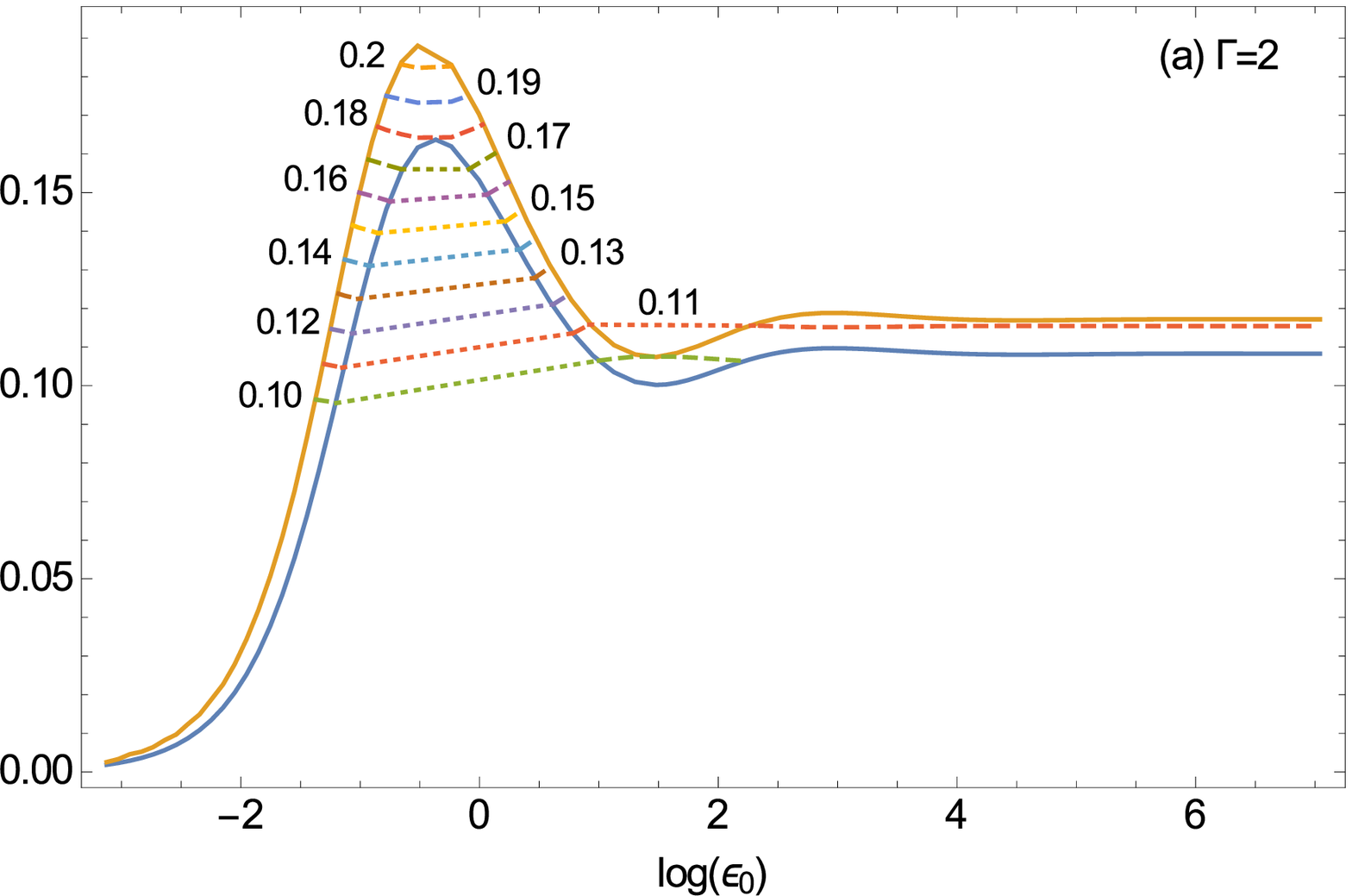}
\includegraphics[scale=0.45]{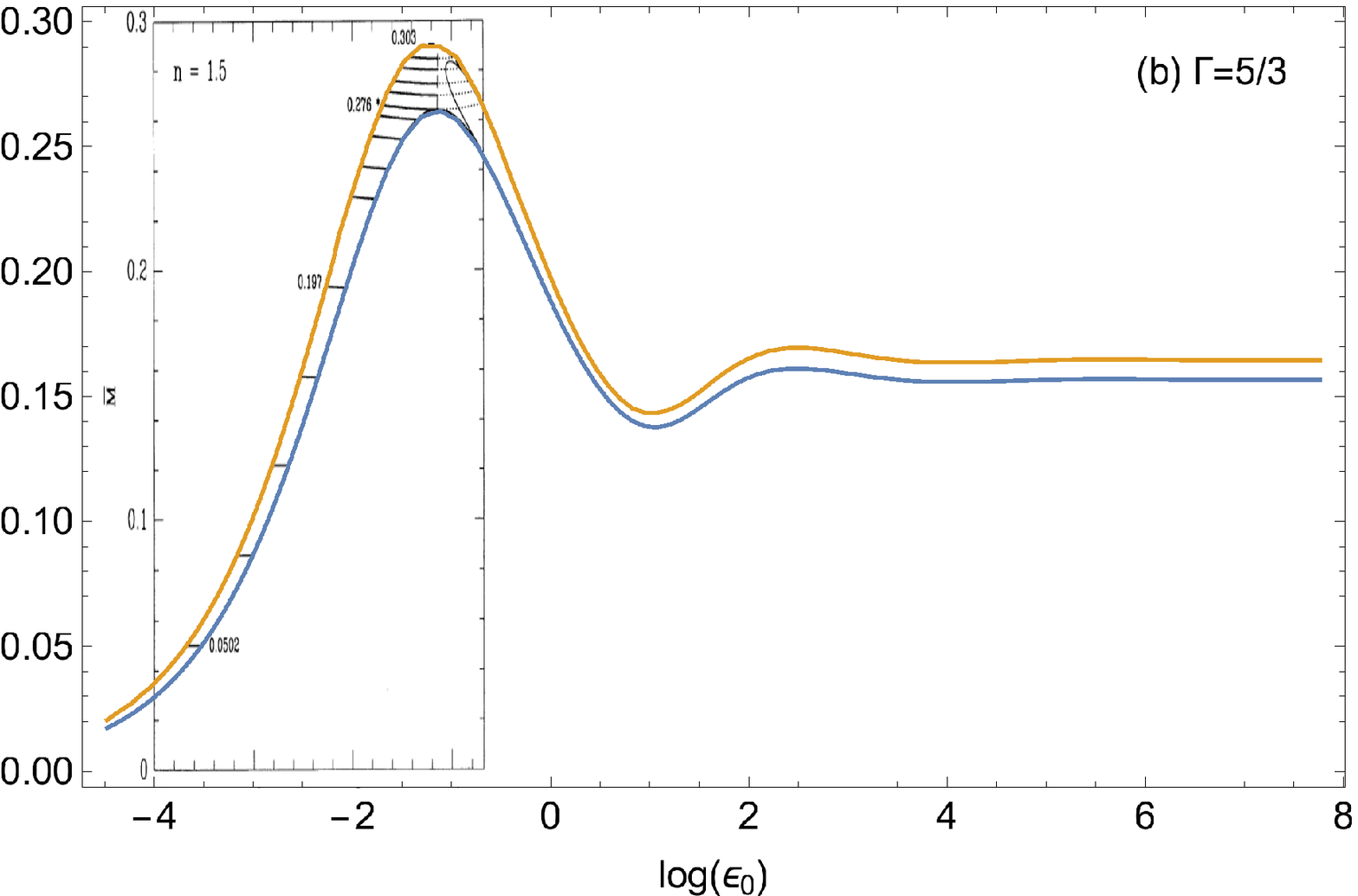}
\includegraphics[scale=0.45]{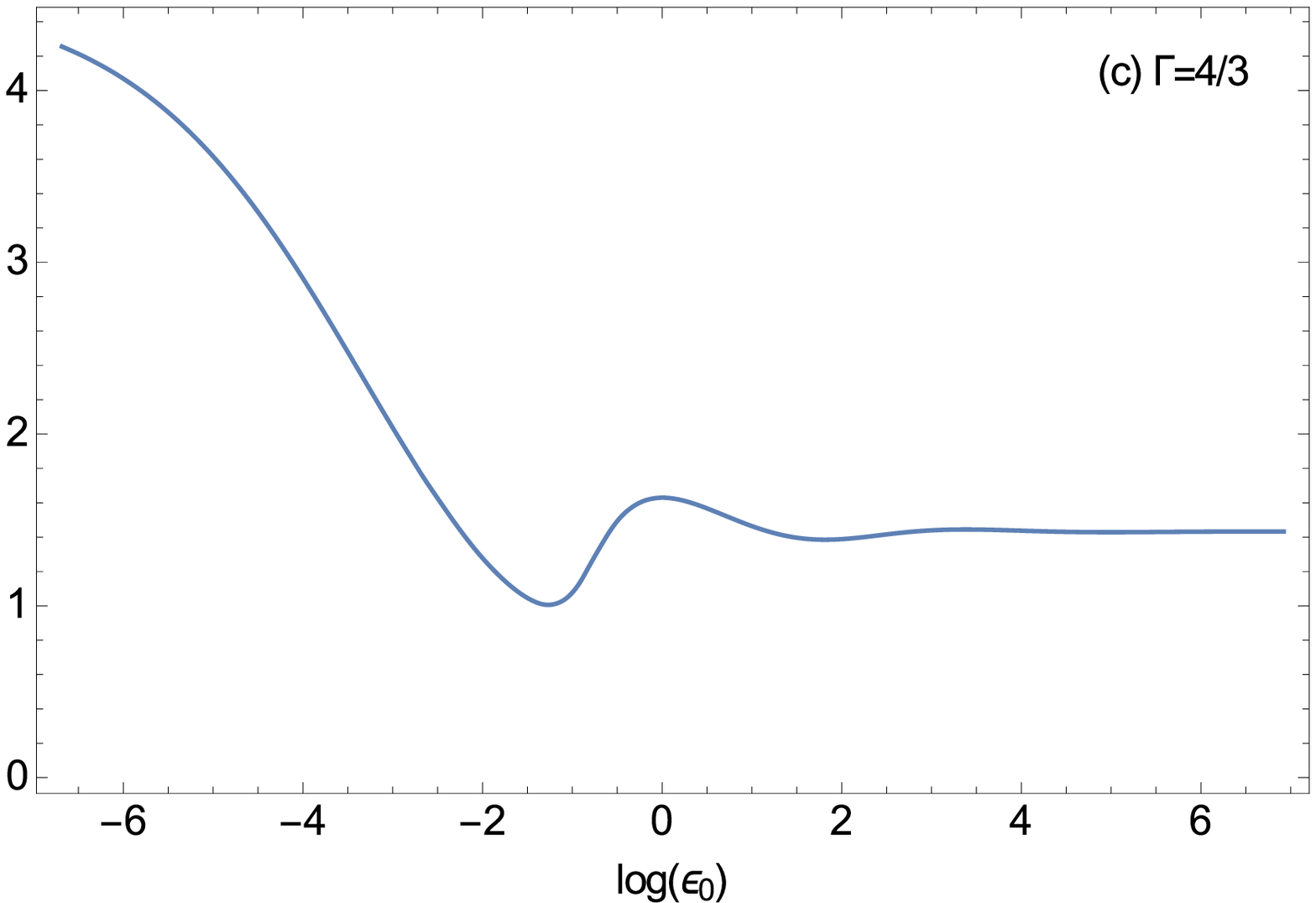}
\end{center}
\caption{\label{fig:mPlots} Solid curves are the mass of the distribution ($M$) as a function of central energy density. The lower solid curve is the spherical (non-spinning) solution. The upper solid curve in (a) and (b) is the maximum mass at the mass-shedding limit (maximum $\Omega_0$); in (c) $\Omega_0$ where mass-shedding occurred was small enough, our simulations did not capture it effectively, so the maximally rotating curve is not shown. In (a) $\Gamma=2$: the dashed curves are contours of constant baryon mass (rest-mass of the constituent particles if they were dispersed to infinity), and the dotted curves are meant only to guide the eye to connect curves with the same baryon mass (the solution space is only between the lower and upper solid curves). In (b) $\Gamma=5/3$, with a stretched trace of Fig.~10 from Ref.~\cite{cook1994} overlayed, showing good agreement between our results and those from Ref.~\cite{cook1994}. In (c) $\Gamma=4/3$.}
}
\end{figure}
\begin{figure}
{
\begin{center}
\includegraphics[scale=0.6]{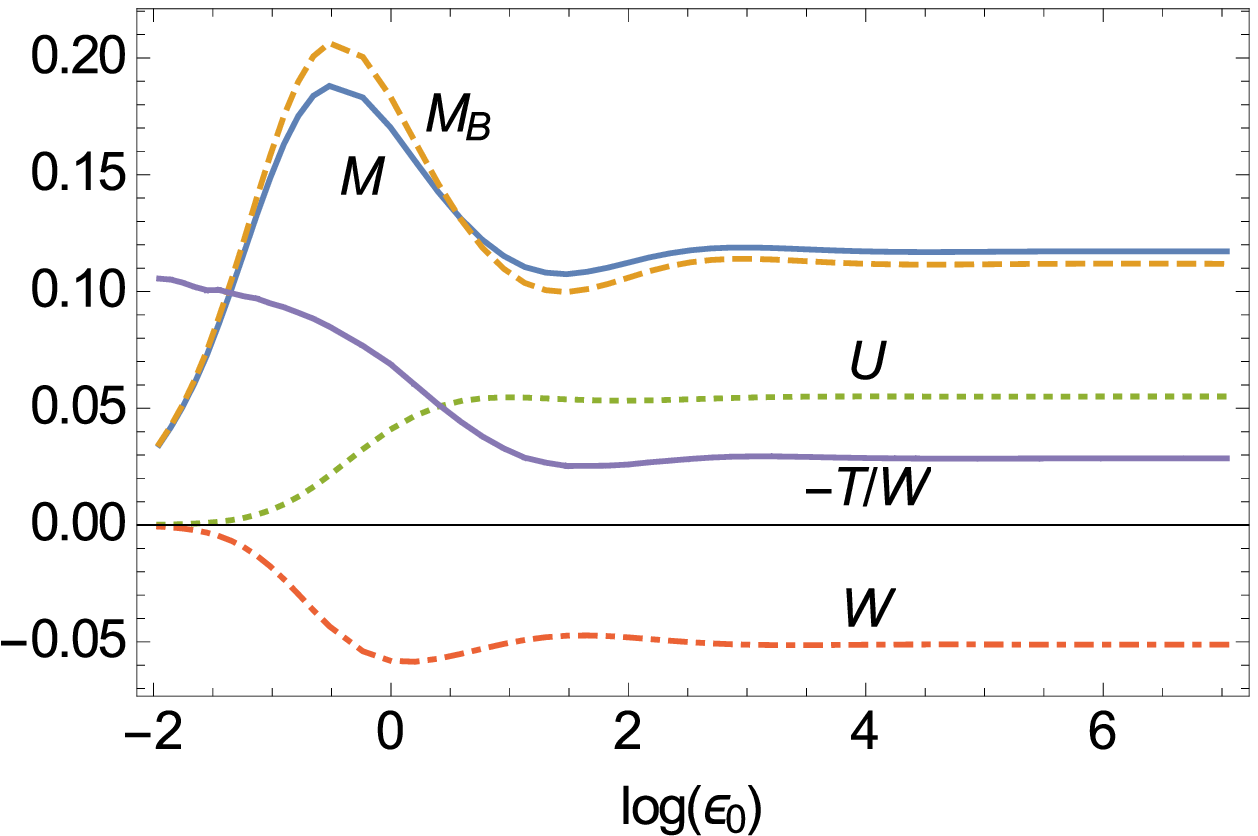}
\end{center}
\caption{\label{fig:mBreakdown}Breakdown of contributions to the total mass of the distribution for $\Gamma=2$ and $\Omega_0$ at its maximum value: $M_B$ is the baryon mass, $U$ is the internal energy (integrated difference between $\epsilon$ and $\rho$), $T$ is the kinetic energy, and $W$ is the gravitational binding energy; $M=M_B+U+T+W$. See Ref.~\cite{stergioulas2003} for explicit definitions of these quantities.}
}
\end{figure}

The mass, $M$, of solutions as a function of central energy density, $\epsilon_0$, is shown in Fig.~\ref{fig:mPlots}. Three adiabatic exponents are shown: $\Gamma=2$, $\Gamma=5/3$, $\Gamma=4/3$. The dashed curves in Fig.~\ref{fig:mPlots}(a) are curves of constant baryon mass (the mass if the particles were at rest with no gravity). 
Note how the contours of constant baryon mass on the right side of the crest of Fig.~\ref{fig:mPlots}(a) are higher than on the left. In fact, the baryon mass is smaller than the total mass for the denser configurations. For $\Gamma=2$ in the case of maximum uniform rotation, the different contributions to the total mass are shown in Fig.~\ref{fig:mBreakdown}. See Ref.~\cite{stergioulas2003} for explicit definitions of these quantities. Also note that all solutions are stable against bar-mode formation, as $-T/W<0.11$ for all solutions, and it decreases with increasing central energy density. 

From Fig.~\ref{fig:trendNearOrigin}, as $H_0$ is increased, for small $s$, the parameters approach a linear function of $s$ near the origin (the dashed lines), before flattening very near the origin (to satisfy the regularity condition). Also, $\gamma$ and $\zeta$ approach the same value as $r\rightarrow 0$, even for the maximally rotating solutions, which means the coordinates approach those of spherical symmetry near the origin; this is a requirement of local flatness at the origin\cite{bonazzola1993}, and is physically reasonable since near the origin the linear velocity due to rotation must approach zero. Therefore, we may use the Tolman-Oppenheimer-Volkoff equation\cite{oppenheimer1939} to study the limiting behavior of the energy density near the origin of the rotating solutions:
\begin{equation}
\begin{array}{rcl}
\frac{dp}{dr_c}&=&-\frac{p+\epsilon}{r_c(r_c-2m)}\left(4\pi p r_c^3+m\right)\\
\frac{dm}{dr_c}&=&4\pi \epsilon r_c^2
\end{array}
\label{tov}
\end{equation}
where $r_c=r e^{\gamma-\nu}$, is the circumferential radius (or the radius associated with the radial coordinates of Ref.~\cite{oppenheimer1939}), and $m$ is the mass contained within a radius $r_c$ (at least $m$ is the total mass when evaluated at the outer radius of a spherical distribution).

In the case of the central energy density being very large, the rest energy becomes negligible compared to the total energy of the fluid, and we may say $p$ and $\epsilon$ are proportional to one another by
\begin{equation}
\begin{array}{rcl}
p&\approx&k_p \epsilon \\
k_p&\equiv&\Gamma-1.
\end{array}
\label{epropp}
\end{equation}
Assuming $\epsilon$ (and $p$) may be written as a power series in $r_c$, we can determine the small $r_c$ behavior given the equation of state, Eq.~\ref{epropp}. Assuming only the lowest order term of the energy series contributes significantly near the origin, approximate $\epsilon$ as
\begin{equation}
\begin{array}{rcl}
\epsilon&\approx& \epsilon_c r_c^q,
\label{approxE}
\end{array}
\end{equation}
 where $q$ is the lowest power in the power series, and $\epsilon_c$ is a constant. Inserting this into Eq.~\ref{tov} yields
\begin{equation}
\begin{array}{rcl}
q k_p \epsilon_c r_c^{q-1}&=&-\frac{1+k_p}{r_c(r_c-2m)}\epsilon_c r_c^q\left(4\pi k_p \epsilon_c r_c^{q+3}+m\right)\\
\frac{dm}{dr_c}&=&4\pi \epsilon_c r_c^{q+2}.
\end{array}
\label{tov2}
\end{equation}
Integrating to find $m(r_c)$ (and noting $m(r_c=0)$ must be zero\cite{oppenheimer1939}), we find in order for Eq.~\ref{tov2} to be consistent, $q$ must be $-2$, and $\epsilon_c$ must satisfy 
\begin{equation}
\begin{array}{rcl}









\epsilon_c&=&\frac{k_p}{2\pi\left(k_p^2 + 6 k_p + 1\right)}\\
\epsilon_c&=&\frac{\left(\Gamma-1\right)}{2\pi(\Gamma^2+4\Gamma-4)}.
\end{array}
\label{p0Constraint}
\end{equation}
In radial coordinates, the spherically symmetric metric may be written as\cite{oppenheimer1939}:
\begin{equation}
\begin{array}{rcl}
ds^2&=&-e^{\nu_c}d t^2+e^{\lambda_c}d r_c^2+r_c^2(d\theta^2+\sin^2\theta d\phi^2),
\end{array}
\label{radialMetric}
\end{equation}
where $\nu_c$ and $\lambda_c$ are convenient metric parameters used in Ref.~\cite{oppenheimer1939}. $\nu_c$ and $\lambda_c$ are determined by the differential-algebraic equations\cite{oppenheimer1939}
\begin{equation}
\begin{array}{rcl}
e^{-\lambda_c}&=&1-2 m/r_c\\
\frac{d\nu_c}{d r_c}&=&-\frac{2}{p+\epsilon}\frac{dp}{dr_c}.
\end{array}
\label{radialMetricEqns}
\end{equation}
Solving these equations using Eqs.~\ref{epropp}, \ref{approxE}, and \ref{p0Constraint} with $q=-2$ gives
\begin{equation}
\begin{array}{rcl}
e^{\lambda_c}&=&\frac{1}{1-8\pi \epsilon_c}\\
e^{\nu_c}&=&\left(\frac{r_c}{r_{c0}}\right)^{\frac{4 k_p}{1+k_p}}=\left(\frac{r_c}{r_{c0}}\right)^{\frac{4(\Gamma-1)}{\Gamma}},
\end{array}
\label{radialMetricValues}
\end{equation}
where $r_{c0}$ is an integration constant with units of length, which is set by the boundary condition at $r_c=\infty$ (e.g. asymptotic flatness or connecting to an appropriate external metric).

The quasi-isotropic coordinates of our spinning solutions limit to isotropic coordinates (not radial coordinates) in the case of spherical symmetry. Therefore, we must convert Eq.~\ref{radialMetricValues} to isotropic coordinates to compare the limiting behavior of our spinning solutions. Using the condition on isotropic coordinates that $g_{\theta\theta}=r^2g_{r r}$, we require a coordinate transformation such that
\begin{equation}
\begin{array}{rcl}
r_c^2&=&r^2e^{\lambda_c}\left(\frac{dr_c}{dr}\right)^2,
\end{array}
\label{radialToIso}
\end{equation}
where $r$ is again our radial coordinate from Eq.~\ref{bonG} (assuming spherical symmetry). Solving this equation yields the following coordinate transformation:
\begin{equation}
\begin{array}{rcl}
r_c&=&r_{c0}\left(\frac{r}{r_{c0}}\right)^{k_r},
\end{array}
\label{coordTransform}
\end{equation}
where we've introduced a new constant $k_r^2\equiv e^{\lambda_c}$; also, an arbitrary integration constant was set in terms of $r_{c0}$ in order to keep both $r$ and $r_c$ with units of length. Using this transformation yields the predicted analytic behavior for our metric potentials near the origin for very large central energy density:
\begin{equation}
\begin{array}{rcl}
\nu=\frac{2 k_p}{k_r(1+k_p)}\ln\left(\frac{r}{r_{c0}}\right)
\end{array}
\label{nuLargeE}
\end{equation}
\begin{equation}
\begin{array}{rcl}
\zeta=\gamma=\left(\frac{3 k_p+1}{k_r(1+k_p)}-1\right)\ln\left(\frac{r}{r_{c0}}\right)
\end{array}
\label{zetaLargeE}
\end{equation}
\begin{equation}
\begin{array}{rcl}
\epsilon=\frac{\epsilon_c}{r_{c0}^2}\left(\frac{r}{r_{c0}}\right)^{-2 k_r}.
\end{array}
\label{eLargeE}
\end{equation}
These analytic expressions are plotted as the dashed lines in Fig.~\ref{fig:trendNearOrigin} confirming this limiting behavior as the central energy becomes large. It also serves as strong support for the numerical models, since the numerical model agrees well with the analytic solution.

In the non-rotating case, we wish to investigate the space of allowed adiabatic exponents. The speed of sound is determined by adiabatically differentiating $p$ with respect to $\epsilon$, $v_c^2=dp/d\epsilon$\cite{tooper1964}. For our equation of state, this yields
\begin{equation}
\begin{array}{rcl}
v_c^2&=&\frac{\Gamma n^{\Gamma-1}}{K_2+\frac{\Gamma}{\Gamma-1}n^{\Gamma-1}},
\end{array}
\label{vSound}
\end{equation}
and in the limit of large $n$ (large $\epsilon$), this limits to 
\begin{equation}
\begin{array}{rcl}
v_c^2&=&\Gamma-1.
\end{array}
\label{vSoundLargeE}
\end{equation}
If $\Gamma>2$, at some large energy density, the speed of sound exceeds the speed of light; therefore, such fluids cannot support physically meaningful solutions at very high energy densities. 

\begin{figure}
{
\begin{center}
\includegraphics[scale=0.45]{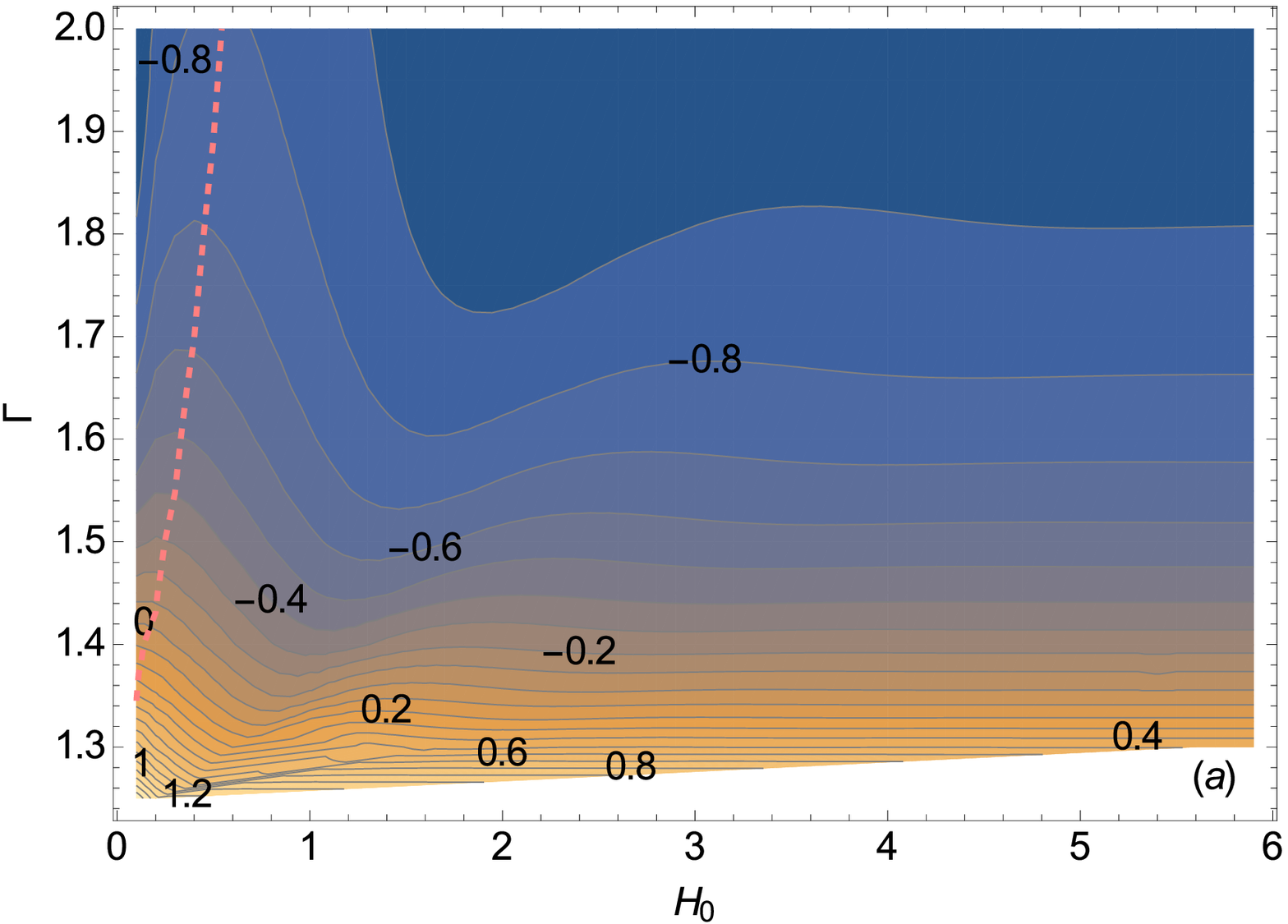}
\includegraphics[scale=0.45]{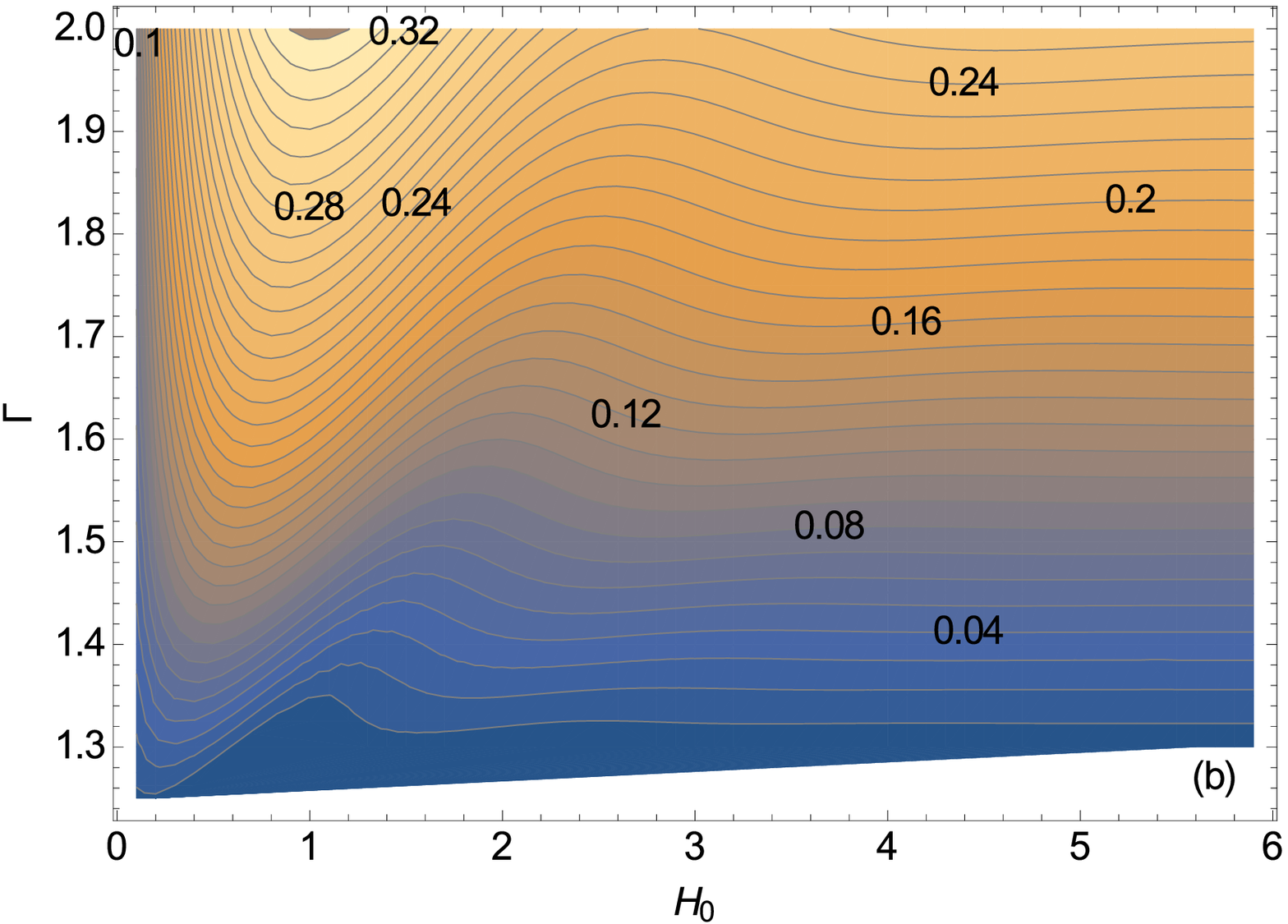}
\end{center}
\caption{\label{fig:gammaH}(a) $\log(M)$ as a function of $H_0$ and $\Gamma$. The jagged lower edge of the plot is where the solver was no longer able to find solutions. The red dotted line is the boundary of stability: solutions to the right of the red dotted line are unstable against radial perturbations. (b) $-(H_0+\nu_0)$ as a function $H_0$ and $\Gamma$. This quantity approaching zero is an indication of the loss of gravity's ability to bind the mass.}
}
\end{figure}

Therefore, we simulated spherically symmetric situations with central enthalpy ranging from $H_0=0.1$ to $H_0=6$, and $\Gamma\leq 2$. $\log(M)$ as a function of $H_0$ and $\Gamma$ is shown in Fig.~\ref{fig:gammaH}(a). The jagged boundary at the bottom of Fig.~\ref{fig:gammaH}(a) is where the solver stopped due to difficulty in finding solutions.

From a mathematical perspective, one way to measure how well gravity binds the matter is the quantity $-(H_0+\nu_0)$, where $\nu_0$ is the central value of $\nu$. If this quantity is less than zero, then from Eq.~\ref{firstIntegral}, $H$ will not approach zero anywhere, the mass would be unbounded, and there is no gravitationally bound solution. This quantity as a function of $H_0$ and $\Gamma$ is shown in Fig.~\ref{fig:gammaH}(b). One may use this to approximate where solutions cease to exist, i.e. where $-(H_0+\nu_0)\rightarrow 0$. As this quantity approached zero, our solver had more difficulty (more iterations, longer time) in converging to a solution, until no solution was found.

As another check of our solutions, in the spherical case, we may use the sufficient condition for dynamic instability from Ref.~\cite{chandrasekhar1964} (if the RHS of Eq.~61 from Ref.~\cite{chandrasekhar1964} is less than zero, the solution is unstable). One must choose a test function, $\xi$, which describes the perturbation of the fluid from equilibrium. This test function must satisfy the condition that at $r_c=0$, $\xi=0$, and the change in pressure due to the displacement at the edge of the distribution is zero. For any test function that satisfies these conditions, if the condition of Ref.~\cite{chandrasekhar1964} is less than zero, then the distribution will be unstable to small perturbations. We used two test functions: $\xi=r_c$, and $\xi=1-\exp(4 r_c)$, which yielded basically indistinguishable boundaries of stability, and which coincided with the first maximum in $M(\epsilon_0)$. This is shown as the red dotted line in Fig.~\ref{fig:gammaH}(a): to the right of the red dotted line, static solutions are unstable to radial perturbations. This is consistent with the results of Ref.~\cite{misner1964}; this consistency again lends confidence to the numerical approach.

\section{Discussion}

We developed a numerical process for calculating uniformly rotating solutions in general relativity, which can solve situations where the densities approach singularity. We compared the behavior of the numerical solutions to the predicted analytical behavior, and found good agreement. We also found good agreement between our solutions and solutions available in the literature. 

These methods can easily be extended to differentially rotating systems, which might show more interesting interplay between the rotation and the singular central mass density.

We studied the stability of the spherical solutions, where we found the first maximum in $M(\epsilon_0)$ marks the boundary between stability and instability. This is also consistent with the literature, and supports our numerical models.

For solutions very close to singular (with very large central energy density), the time-time component of the metric approaches zero. Therefore, the evolution near the singularity, although unstable, might appear somewhat stable on timescales which are reasonable in astrophysical phenomena. Additionally, the mass of the more singular distributions is greater than the baryon mass; this means, in principle, these distributions may just as likely explode as collapse\cite{gourgoulhon1991}. In any case, time dependent simulations starting with a solution which is nearly singular could be interesting.

Since the solutions presented here can provide consistent initial conditions arbitrarily close to having a singular central energy density (and infinite curvature), using them as initial data in time dependent simulations could also be interesting in the study of black-hole formation.



\end{document}